\documentclass[12pt]{article}
\usepackage{amsmath,amssymb,graphicx,mathrsfs,hyperref,slashed,setspace}

\usepackage{float}
\usepackage{latexsym}
\usepackage{bm}
\usepackage{blindtext}
\hypersetup{
	colorlinks=true,
	linkcolor=blue,
	filecolor=black,
	urlcolor=black,
	citecolor=blue,
}

\newcommand{\hide}[1]{}

\newcommand{\be}{\begin{equation}}
\newcommand{\ee}{\end{equation}}
\newcommand{\bea}{\begin{eqnarray}}
\newcommand{\eea}{\end{eqnarray}}

\def\({\left(} \def\){\right)}

\begin{document}
\title{\vspace{-1.8in}
{Frozen stars: Black hole mimickers sourced by a string fluid}}
\author{\large Ram Brustein${}^{(1)}$,  A.J.M. Medved${}^{(2,3)}$
\\
\vspace{-.5in} \hspace{-1.5in} \vbox{
\begin{flushleft}
 $^{\textrm{\normalsize
(1)\ Department of Physics, Ben-Gurion University,
   Beer-Sheva 84105, Israel}}$
$^{\textrm{\normalsize (2)\ Department of Physics \& Electronics, Rhodes University,
 Grahamstown 6140, South Africa}}$
$^{\textrm{\normalsize (3)\ National Institute for Theoretical Physics (NITheP), Western Cape 7602,
South Africa}}$
\\ \small \hspace{0.57in}
   ramyb@bgu.ac.il,\  j.medved@ru.ac.za
\end{flushleft}
}}
\date{}
\maketitle

\begin{abstract}

The frozen star is a non-singular, ultracompact object that, to an external observer, looks exactly like a Schwarzschild black hole, but with a different interior geometry and matter composition.  The frozen star needs to be sourced by an extremely anisotropic fluid, for which the sum of the radial pressure and energy density is either vanishing or  perturbatively small.   Here, we show that this matter can be identified with the string fluid resulting from the decay of an unstable $D$-brane or a brane-antibrane system at the end of open-string tachyon condensation. The string fluid corresponds to flux tubes emanating from the center and ending at the Schwarzschild radius of the star. The effective Lagrangian for this fluid can be recast into a Born-Infeld form. When the fluid Lagrangian is coupled to that of Einstein's Gravity, the static, spherically symmetric solutions of the equations of motion are shown to be the same as those describing the frozen star model. Frozen stars can therefore be viewed as gravitationally back-reacted BIons. The Born-Infeld Lagrangian provides a complete set of equations that describe the dynamics of the frozen star in a generic state, which is not necessarily static nor spherically symmetric. Additionally, this description provides a new physical perspective on the structure of the frozen star in terms of the corresponding electric fields and charges. The electric field is sourced by a point-like charge at the center of the star, while its  outer layer is equal and oppositely charged. The electric force between the charges is offset because the mass of the star is fixed.
\end{abstract}
\newpage
\setstretch{1.5}

\renewcommand{\baselinestretch}{1.5}

\section{Introduction}

The frozen star   is a type of  black hole (BH) mimicker: a static, spherically symmetric solution of Einstein's equations that is a regular and horizonless alternative to  the singular
Schwarzschild  solution \cite{bookdill,BHfollies,popstar,trajectory,fluctuations}. Furthermore, the frozen star is able to reproduce all of the standard thermodynamic properties of  a Schwarzschild BH of the same mass \cite{U4Euclidean}.
 It is also possible to incorporate rotation into the frozen star model \cite{notstevekerr}, resulting in a regular and horizonless mimicker of the Kerr solution.

The motivation for the interior metric~(\ref{linement}) follows from the initial impetus for the model itself. As discussed at length in \cite{bookdill} (also see \cite{BHfollies}), this solution  is  meant to be an effective classical description of a highly quantum state
for the object's interior that is known as the polymer model \cite{strungout,emerge}. In this picture, the BH mimicker consists of an extremely hot collection of long, closed, interacting, fundamental strings. One can view the polymer model as the microscopic description  of the frozen star. As the story unfolds, this string picture will  come around full circle, as we arrive at yet another  description of the interior that is  motivated by string theory but differently from the polymer model.

We will show that the frozen star is sourced by a fluid of cold strings resulting from the decay of an unstable $D$-brane or a brane--antibrane system at the end of open-string tachyon condensation \cite{Sen,Sen2,Sen3,Sen4,Sen5}. Gibbons, Hori and Yi (GHY) \cite{GHY} (also see \cite{Yi}) reformulated  Sen's  effective action describing the state of unstable $D$-branes after the process of tachyon condensation. Many more references to discussions on this process can be found in \cite{SenReview}. The reformulated Lagrangian is of a specific Born--Infeld form which describes a  fluid of rigid electric-flux tubes. In \cite{Yi}, the possible bending and stretching of the flux tubes were also considered. As was duly noted in \cite{GHY}, a similar form of Lagrangian with a two-form field first appeared in \cite{NO}. After that,  the same Lagrangian was proposed in the context of the cloud-of-strings model in \cite{Letel} and the string-dust model in \cite{Stach} and was discussed in a cosmological context as a  ``hedgehog compactification'' in   \cite{hedge1,hedge2}.

When gravity is neglected, the Born--Infeld theory of $Dp$-branes has been shown to give rise to  spherically symmetric, static, solitonic solutions of finite energy that are known as BIons \cite{GibbonsBIons}. (See \cite{GibbonsRev} for a later review.) BIons were shown to have a point-like source in their core with strings, or flux tubes, emanating from the core and going all the way to infinity. From this perspective,  frozen stars can  be viewed as a specific form of gravitationally back-reacted BIons whose energy and spatial extent are both finite.

The Born--Infeld Lagrangian completes the Einstein equations and describes the dynamics of the frozen star in a generic state, which is not necessarily static nor spherically symmetric. We can thus show that the frozen star solution is a consistent and complete model which describes an ultracompact object whose equilibrium state is practically indistinguishable from that of a Schwarzschild BH.

Although beyond the scope of the current investigation, the framework that is developed here
should  also allow   us to study extensions  of our model to rotating frozen stars  \cite{notstevekerr}  and to
the ``defrosted star'' model, which allows for deviations away from $\;\rho+p=0\;$ \cite{fluctuations}. The latter is a necessary step for the frozen star interior to support dynamical modes  because of  the ultrastability of the undeformed model \cite{bookdill,popstar,fluctuations}.
The current framework should  eventually enable us to study extensions  of our model to cases in which the  departures from equilibrium physics are macroscopic --- the importance of which has been stressed in \cite{ridethewave}.   Out-of-equilibrium physics could  prove to  be important in describing  the dynamics of astrophysical BH mergers, which would  help in distinguishing the frozen star from the Schwarzschild solution, as well as  from  other BH mimickers.

\section{The frozen star}

The simplest form of frozen star metric, which is the one that will be considered in this paper, is as follows:
\be
ds^2\; =\; -\varepsilon^2 dt^2 + \frac{1}{\varepsilon^2} dr^2 + r^2 (d\theta^2+ \sin^2 \theta d\phi^2)\;,
\label{linement}
\ee
where  $\varepsilon^2$ is a small, constant and  dimensionless parameter.~\footnote{The same parameter was referred to as
$\epsilon$ or $\varepsilon$ in our earlier  papers on this topic.}
A recent paper that  used data from the Event Horizon Telescope constrained the parameter to be extremely small,
$\;\varepsilon^2\lesssim 10^{-22}\;$ \cite{clowncar}. The outermost surface of the star is pushed out slightly from the location of the would-be horizon, $\;R\sim 2MG(1+\varepsilon^2)\;$, where
$R$ is the star's radius and $M$ is its mass. Additionally, just like the horizon-like outer surface,  each radial slice of the interior is a surface of exponentially large but finite redshift and thus  can be viewed, approximately, as a marginally trapped surface.

The stress--energy tensor $T^a_{\;\;b}$ that is needed to source this simplest frozen star geometry  is distinguished by having a radial component of pressure, $\;p\equiv p_r\;$, that
takes on the most negative value that is allowed by causality: $\;p=-\rho\;$, where $\rho$ is the energy density. Meanwhile, the  transverse components of pressure $p_{\perp}\;$ are vanishing,
\bea
\rho&=& -T^t_{\;\;t}\;=\; -\frac{1}{8\pi G}G^t_{\;\;t}\;=\; \frac{1}{8\pi G} \frac{1-\varepsilon^2}{r^2}\;, \label{rho} \\
p &=& T^r_{\;\;r}\;=\;\frac{1}{8\pi G}G^r_{\;\;r}\;=\; -\frac{1}{8\pi G} \frac{1-\varepsilon^2}{r^2}\;, \label{pee} \\
p_{\perp} &=& T^{\theta}_{\;\;\theta}\;=\; T^{\phi}_{\;\;\phi}\;=\;
\frac{1}{8\pi G}G^{\theta}_{\;\;\theta}  \;=\;
0\;.
\label{que}
\eea
It follows that
\be
2GM \;=\; 2G\int^R dr\;4\pi r^2 \rho\;=\; R(1-\varepsilon^2)\;.
\ee

The frozen star geometry can be viewed as a spherically symmetric collection of straight, radially pointing, rigid strings ({\em i.e.}, a ``hedgehog''),  each with a constant tension of $ \frac{1}{8\pi G}$ \cite{trajectory}.  The geometry  is mildly singular near the center of the star, as are $\rho$ and $p$, so that  a very small concentric sphere must be regularized to
ensure that these densities remain finite. This process was described  in
\cite{trajectory}. Also, as detailed in \cite{popstar}, a matching process is required near the outermost layer of the star so that the metric in Eq.~({\ref{linement}) and its corresponding stress tensor in Eqs.~(\ref{rho}-\ref{que})  match smoothly to the exterior Schwarzschild geometry. Later in the paper, we will offer a different perspective on the regularization and smoothing in terms of a point-like charge at the center of the star and an equal and  opposite charge that is  distributed uniformly over the surface of the star.

A perk  of the maximally negative radial pressure is that  the frozen star model is able to
evade the  singularity theorems of Hawking and Penrose \cite{PenHawk1,PenHawk2}. For a finite value of $\varepsilon^2$, a trapped surface is never actually formed and having $\;p+\rho=0\;$  means that geodesics do not converge. The way to understand this is to realize that the equation of state $\;p+\rho=0\;$ can also be viewed as the saturation of the radial component of the null-energy condition. Thus, the conditions under which the singularity theorems are valid are not satisfied by the frozen star geometry.  Similarly, a frozen star evades the ``Buchdahl-like'' bounds which limit the compactness of matter \cite{Buchdahl,chand1,chand2,bondi,MM} by having a large negative pressure throughout its interior.

An important characteristic of the frozen star geometry is that the deviations from the Schwarzschild solution are large throughout the interior of the object; that is, on horizon-length scales. This goes against common lore that singularity resolution requires  some quantum  corrections only near the would-be singularity, but such reasoning has been shown to lead to  energy loss by radiation  that far exceeds  the original mass of the object \cite{frolov,visser}.

 \section{Born-Infeld effective Lagrangian and BIons}

Here, we review for completeness  relevant portions of the analysis of Gibbons, Hori and Yi (GHY) in \cite{GHY} (also see \cite{Yi}), where further details and references can be found. Our conventions are that an index of  0 denotes  time, one of $a$,$b$,$\cdots$ denotes an arbitrary spacetime dimension   and  one of $i$,$j$,$\cdots$ denotes an arbitrary  spatial
dimension. We assume three large spatial dimensions for concreteness and, for spherical coordinates, $\;(0,1,2,3)=(t,r,\theta,\phi)\;$.

The starting point for the analysis in  \cite{GHY}  is  Sen's effective action for the decay of unstable  $D$-branes;  specifically, at the end of  tachyon condensation and near the  minimum  of the tachyon potential, where it is vanishing \cite{Sen,Sen2,Sen3,Sen4,Sen5}. Sen's effective Lagrangian (density) can be expressed as
\be
{\cal L}\;=\;- V(T)\sqrt{-Det\left(\eta+ 2\pi \alpha'{\cal F}\right)}\;+\;
\sqrt{-\eta}
A^a J_a\;,
\label{LofT}
\ee
where $V(T)$ is the tachyon potential or, equivalently, the $D$-brane  tension,
$2\pi \alpha'$ is the inverse of the fundamental string tension,
$\;\eta^{a}_{\;\;b}=\delta^{a}_{\;\;b}\;$ is the Minkowski background metric,
$\;{\cal F}_{ab}=\partial_{a}A_{b}-\partial_{b}A_{a}\;$ is the field-strength  tensor for  the gauge field $A_a$ and $J_a$ is the source, a 4-vector current density. We have not included the kinetic terms  for the tachyon field and for  the scalars that  are associated with the additional transverse dimensions. These fields are readily restored if need be (see \cite{Yi} for formal  details) but are not necessary for the current analysis.

The Lagrangian (\ref{LofT}) vanishes when $V(T)$ vanishes; however, the Hamiltonian ${\cal H}$ is  well defined.
In the case of no magnetic sources, this is
\be
\;{\cal H}\;=\;D^i{\cal F}_{0i}-{\cal L}\;=\;E_i D^i\;,
\label{hamil}
\ee
where the last equality has used  the fact that  $\;{\cal L}=0\;$ at the minimum of
the potential,
$\;E_i={\cal F}_{0i}\;$ is the electric field and
 $\;D^i$ is the electric displacement which, in the current case, is the canonical conjugate of the gauge field,  $\;D^i=\frac{\delta {\cal L}}{\delta (\partial_0  A_i)}\;$.
 The displacement $D^i$ is naturally preferred over  $E^i$ to play the role of the ``electric field'' because it is the field  in a Born--Infeld theory that always satisfies the Gauss'-law constraint,
\be
\partial_i D^i\;=\;J_0\;=\;\rho_e\;.
\label{sourceD}
\ee

More generally, the  Hamiltonian can  be expressed as
\be
{\cal H}\;=\;\frac{\delta {\cal L}}{\delta  (\partial_0 A_i)}-{\cal L}\;=\;\frac{1}{2\pi\alpha'}\sqrt{D^i D_i + {\cal P}^i{\cal P}_{i}}\;,
\ee
where $\;{\cal P}_i=-{\cal F}_{ij} D^{j}=\left(\vec{D}\times\vec{B}\right)_i\;$ is the conserved momentum associated with spatial translations.  The definition of the magnetic induction is standard, $\;B^i=\frac{1}{2}\epsilon^{ijk} {\cal F}_{jk}\;$.

The sources can be included implicitly by imposing the Gauss'-law constraint~(\ref{sourceD}),
Ampere's law $\;\partial_i {\cal F}^{i}_{\;\;j}- \partial_0{E}_j= J_j\;$,
along with  the Bianchi identities, which include Faraday's law $\;{\nabla \times}\;\vec{E}+\partial_0{\vec{B}}=0\;$ and
\be
\partial_i B^i\;=\;0\;.
\label{sourceBD}
\ee
The last equation assumes that there are no sources for magnetic monopoles.  The inline equations above Eq.~(\ref{sourceBD}) are of less importance here, as our main concern is the case for which magnetic sources and  time-dependent fields are absent.

To obtain a useful Lagrangian, GHY follow techniques from \cite{Tsey} and regard  the magnetic fields
${\cal F}_{ij}$ as the conjugates with respect to a new set of  dual variables $K_{ij}$ such that $\;K^{ij}=2\frac{\delta {\cal H}} {\delta {\cal F}_{ij}}$. The fields ${\cal F}_{ab}$ and  ${\cal K}_{ab}$ --- the latter being
an extension of $K_{ij}$ as defined in Eq.~(\ref{KalK})   ---
should be regarded as independent variables. This will be relevant when we derive the equations of motion (EOM) for the tensor field ${\cal K}$.

The resulting Lagrangian is then defined  by an appropriate  Legendre transformation,
\be
{\cal L}'\;=\; {\cal H}-\frac{1}{2}{\cal F}_{ij}K^{ij}\;=\;\frac{1}{2\pi\alpha'}\sqrt{D^iD_i-\frac{1}{2}K^{ij}K_{ij}}\;,
\label{canon}
\ee
and is clearly non-vanishing in general. The latter equality can be obtained using the relation
\be
 \frac{1}{2}K^{ij}K_{ij}\;=\;\frac{1}{{\cal H}^2}{\cal P}^{i}{\cal P}_{i}D^j D_j\;.
 \label{thekey}
\ee

GHY then define a two-form field ${\cal K}_{ab}=\partial_a \widetilde{A}_b- \partial_b \widetilde{A}_a \;$ that acts as an effective field strength,
\bea
{\cal K}_{0i}&=& D_i
\;, \cr
{\cal K}_{ij}&=& K_{ij}\;.
\label{KalK}
\eea
The new Lagrangian can now be written in a manifestly Born--Infeld form,
\be
{\cal L}'\;=\; \frac{1}{2\pi \alpha'}\sqrt{-\frac{1}{2}{\cal K}^{ab}{\cal K}_{ab}}\;,
\label{newL}
\ee
where the negative sign is a consequence of the time--time component of the metric appearing in the contraction when $\;a=0\;$ and $\;b=i\;$ (and {\em vice versa}). Here, the (effective) electric-field  term  is presumed to be the dominant one.
In the case of no magnetic sources, one finds that the new Lagrangian is the same as the original Hamiltonian,
$\;{\cal L}' \;=\; E_iD^i \;$.

There is a subtlety in this procedure in that Eq.~(\ref{KalK}) implies that $\;{\cal K}\wedge {\cal K}=0\;$ but, given that this constraint is in effect,
the canonical analysis of ${\cal L}'$ does not lead back to the same Hamiltonian ${\cal H}$. As explained  in \cite{GHY}, this
situation can be rectified by adding a Lagrange-multiplier term to ${\cal L}'$ that imposes the constraint explicitly.

To summarize, the Born--Infeld Lagrangian describing the string fluid is the following:
\be
{\cal L}'\;=\; \frac{1}{2\pi \alpha'}\sqrt{-\frac{1}{2}{\cal K}^{ab}{\cal K}_{ab}}\;+\; \lambda_1 \epsilon^{abcd}{\cal K}_{ab}{\cal K}_{cd}\;+\sqrt{-\eta} \;A^aJ_a\;,
\label{lagrangey}
\ee
where $\lambda_1$ is the Lagrange multiplier and the constraint is automatically satisfied (and the constraint term vanishes) by imposing Eq.~(\ref{KalK}). The connection with strings follows from the fact that ${\cal K}^{ab}$ can be identified as a surface-forming bivector \cite{GHY}, which can then be interpreted as a  cross-sectional slice  of the world sheet of an open string or a flux tube.

The EOM for this Lagrangian, $\;d{\cal L}'=0\;$,
are most transparent when expressed in terms of the original field-strength tensor as these are
equivalent to the original  Bianchi identities \cite{GHY},
\be
d{\cal F}\;=\;0\;.
\ee
On the other hand, the  Bianchi identities for the new field-strength tensor
\be
 d{\cal K}\;=\;0\;
 \ee
are equivalent to  the EOM for the original Lagrangian  \cite{GHY}.

There is also a subtlety concerning the source term.  The original  source term in Eq.~(\Ref{LofT}) is unaffected by the Legendre transformation; however, $A_a$ is not the gauge field for the field strength ${\cal K}_{ab}$.
Let us denote its gauge field  as $\widetilde{A}_a$, then  $\;D_i=\partial_i \widetilde{A}^0\;$ whereas $\;E_i=\partial_i A^0\;$.  Fortunately, this tension is resolved because, as we have noted, the two types of field strengths are independent, as must also be true of their respective gauge fields.  It follows that we can  vary the  Lagrangian ${\cal L}'$
with respect to either  gauge field, and it happens to be the variation with respect to
$A^{a}$ that leads to the expected Gauss's-law constraint in Eq.~(\ref{sourceD}).
This is most clear in the case of no magnetic sources, as can be seen from the form of the inline equation
for ${\cal L}'$ below Eq.~(\ref{newL}).

If there are no bulk sources --- our case of particular interest --- the stress--energy tensor is given by
 \bea
 T_{ab} \;=\; 2\frac{\delta {\cal L}'}{\delta \eta^{ab}}
        \;=\; \frac{1}{2\pi\alpha'}\frac{{\cal K}_{a}^{\;\;c}{\cal K}_{bc}}{\sqrt{-\frac{1}{2}{\cal K}^{ab}{\cal K}_{ab}}} \;.
 \label{TBI}
 \eea
That this is the appropriate definition of $T_{ab}$ is justified in \cite{Letel}.
In terms of the effective electric fields,  magnetic fields and momenta,
\bea
T_{00} &=&  \frac{1}{2\pi\alpha'}\frac{D^i D_i}{\sqrt{D^i D_i -\frac{1}{2}K^{ij}K_{ij}}}\;=\; {\cal H}\;, \label{Too} \\
T_{0i}  &=&    \frac{1}{2\pi\alpha'}\frac{D^j K_{ij}}{\sqrt{D^i D_i -\frac{1}{2}K^{ij}K_{ij}}}\;=\; -{\cal P}_i\;,
\label{Toi}\\
T_{ij} &=&  \frac{1}{2\pi\alpha'}\frac{-D^i D_j+K_{i}^{\;\;k}K_{jk}}{\sqrt{\pi^i\pi_i -\frac{1}{2}K^{ij}K_{ij}}}\;=\;
\frac{1}{2\pi\alpha'}\frac{-D_iD_j+{\cal P}_i{\cal P}_j}{{\cal H}}\label{Tij}\;,
\eea
where the right-most relations make use of Eq.~(\ref{thekey}).

The simplest solution of the EOM is a static, spherically symmetric configuration, which happens when $\;{\cal K}_{01}=-{\cal K}_{10}\;$ and all other elements  of this field strength  vanish. In which case, $\;{\cal K}^{ab}{\cal K}_{ab}=-2 D_1 D^1$.  Then the only non-vanishing elements of $T_{ab}$ are
\be
T_{00} \;= -\;T_{11}\; =  \;\frac{1}{2\pi\alpha'}\sqrt{D^1D_1}\;,
\label{SS}
\ee
implying that $\;p=-\rho\;$ and $\;p_\perp=0\;$. This is the so-called  string fluid \cite{GHY}, for which
\be
E_1\;=\;\frac{1}{2\pi \alpha'} \frac{D_1}{\sqrt{(D_1)^2}}\;=\;\frac{1}{2\pi \alpha'}\;,
\label{Er}
\ee
and so
\be
T_{00} \;= \;  E_1 D^1\;.
\label{T00DE}
\ee
Importantly, the bulk portion (or first term) of the Lagrangian~(\ref{lagrangey}) is exactly the same as $T_{00}$,
\be
{\cal L}^{\prime}_{bulk} \;=\;\frac{1}{2\pi\alpha'}\sqrt{D^1D_1} \;=\;  E_1 D^1\;.
\label{bulky}
\ee

The static and spherically symmetric case is also related to the BIon solution \cite{GibbonsBIons}, for which $\;D_1=\frac{q}{4 \pi r^2}\;$, corresponding to a point-like charge at the origin $\;\partial_i D^i= q ~\delta^{(3)}(\vec{r})\;$. BIons can therefore be viewed as being sourced by a fluid of electric flux lines that are emanating radially from a point source at the center  and extending all the way to infinity.

\section{Frozen stars as gravitationally backreacted BIons}

In this section, we couple the BIons to Einstein's gravity, discuss the backreaction on the  BIons and, finally, show that they correspond to solutions of the frozen star model.

The complete gravitational and matter action  includes, in addition to the Einstein--Hilbert and Born--Infeld Lagrangians, the constraint and  source terms
\be
S_{GBI}\;=\;\int d^4 x \left\{ \sqrt{-g}R +\frac{1}{2\pi \alpha'}\sqrt{-\frac{1}{2}{\cal K}^{ab}{\cal K}_{ab}}+\lambda_1 \epsilon^{abcd}{\cal K}_{ab}{\cal K}_{cd} +  \sqrt{-g} J_a A^a \right\} \;.
\ee
As discussed at the end of this section, a second constraint which fixes the mass is still to be included.
But, as a surface term, it will not affect the bulk EOM. We will also be assuming that there are no sources in the bulk.

The gravitational EOM are
$ \;
T^a_{\;\; b}=\frac{1}{8\pi G}G^{a}_{\;\;b} \;
$,
with $T^a_{\;\; b}$ given in  Eq.~(\ref{TBI}), again because there are no bulk sources. The Born--Infeld EOM  are  as discussed in the previous section, except partial derivatives should be replaced by covariant derivatives (although often inconsequential because of the anti-symmetry properties of the field-strength tensor). The same applies to the Bianchi identities and source constraints.

To reproduce the frozen star solution, we will need to choose specific source terms and  impose certain boundary conditions such that  the total charge of the frozen star vanishes  and its total mass is fixed (the latter constraint will be discussed further on).  The source current $J_a$, for which  only $J_0$ is non-vanishing, therefore includes a point-like positive electric charge at the star's center  and a negative charge of equal magnitude which is spread evenly over the outer surface of the star.

One might be concerned that the attractive electric force between the opposite charges in the core and the outer layer will endanger its stability. However, as discussed later, fixing the mass of the star exactly offsets this attractive force. The reason being that a fixed mass for the solution translates into a fixed radius, so that the outer layer cannot move inwards (or outwards) in response to a supplementary  force. This was a key ingredient in the discussion of the stability of the frozen star \cite{bookdill,popstar,fluctuations}.

The frozen star solution can be related to the BIon solution. The matching can be made precise for the static, spherically symmetric case by  using the ${}^t_{\ t}$ component of the Einstein equations along with Eq.~(\ref{SS}),
\be
\rho_{FS}\;=\;\frac{1}{8\pi G}\frac{1-\varepsilon^2}{r^2} \;=\;   \frac{1}{2\pi\alpha'}\sqrt{D^1 D_1}\;,
\label{singlepeak}
\ee
such that $\rho_{FS}$ is the energy density as expressed in Eq.~(\ref{rho}).  The  result is
\be
\sqrt{D^1D_1}\;=\; \frac{ \alpha'}{4 G} \frac{1}{r^2} + {\cal O}[\varepsilon^2]\;,
\label{twinpeaks}
\ee
corresponding to the existence of an electric point-like charge at the center of the star, since the solution
follows from
$
\;\nabla_i D^i = \nabla_1 D^1= 
q_{core} ~ \delta^{(3)}(\vec{r})\;.
$
It can then be deduced that
\be
q_{core} \;=\;\pi\frac{\alpha'}{G} \;.
\ee

An interesting feature of the solution is that the ratio ${ \alpha'}/{G} $ in weakly coupled string theories scales as $\;\sim 1/g_s^2\;$, where $g_s^2$ is the closed string coupling. Moreover,  the  description of the  solution as radial flux lines  fits in well with the hedgehog picture of the frozen star that was recalled in Section~2.

The total charge of the frozen star needs to vanish, as its energy density, and thus its electric field, vanishes in the exterior region  where the geometry is described by the vacuum Schwarzschild solution. Correspondingly, there has to be a charge
$q_{out}$ that is spread out uniformly  over the star's outer surface layer and equal to $-q_{core}$. This leads  to a source term at the outer surface as follows:
\be
\left(J_0\right)_{out}\;=\; \frac{q_{out}}{4\pi r^2} \delta(r-R)\;=\;-\frac{q_{core}}{4\pi r^2} \delta(r-R)\;.
\label{J0Out}
\ee

\begin{figure}[t]
\vspace{-.95in}
\begin{flushright}
\includegraphics[height=8cm]{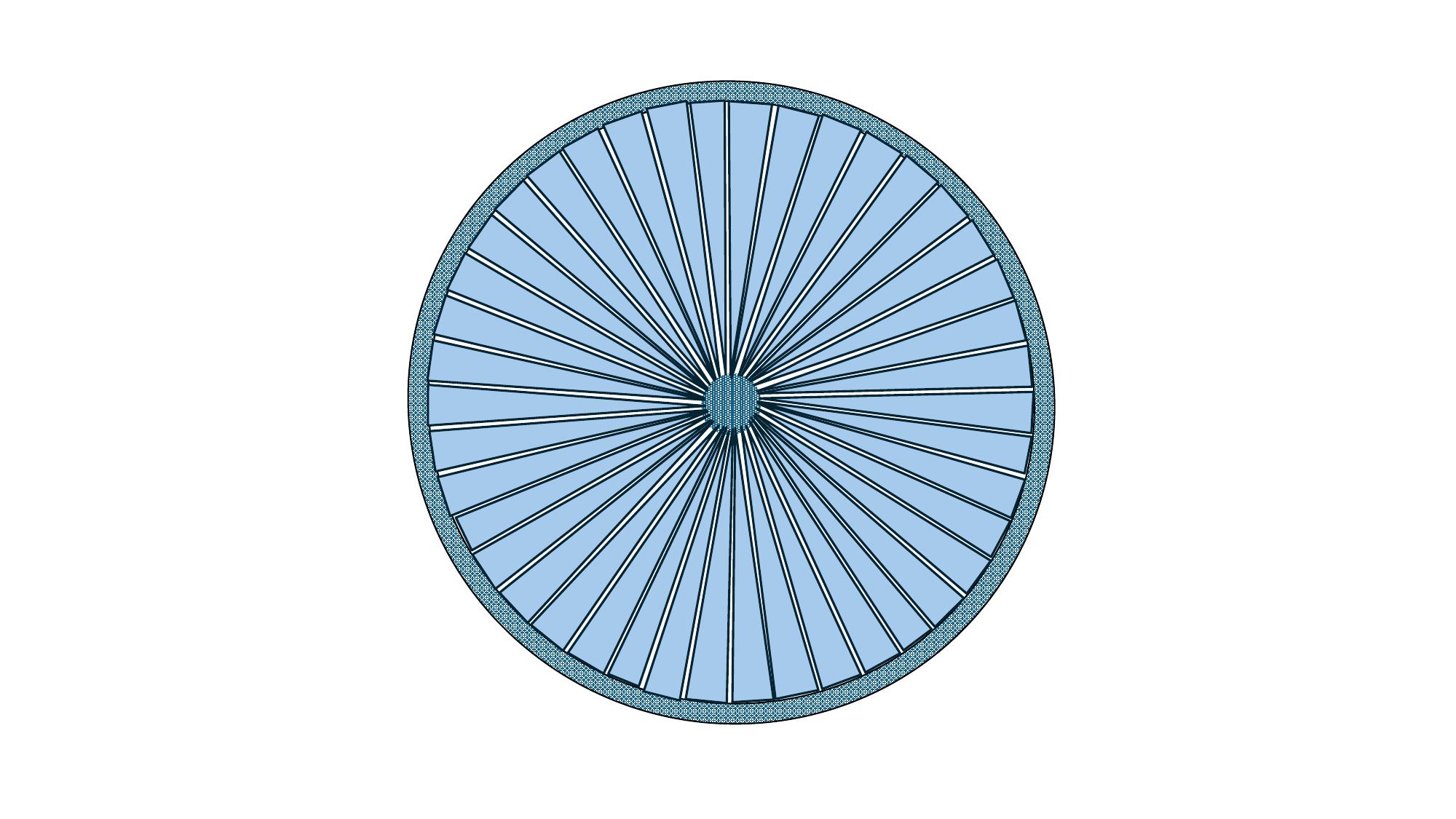}
\end{flushright}
\vspace{-.5in}
	\caption{A frozen star sourced by a fluid of electric flux lines that are emanating radially from a  point-like charge in its core and ending on its outer layer which is oppositely charged. The electric force between the charges is offset because the mass of the star is fixed.}\label{fig}
\end{figure}

Let us now discuss the boundary condition that fixes the mass of the frozen star. This needs to be formally imposed at infinity, as this is the location where  the mass is  defined by way of a surface integral.
However, because the solution outside of the frozen star is strictly a vacuum, the mass can also evaluated as a surface  integral at the outermost radius of the star, or, alternatively, as a volume integral that enforces  the constraint $\;\int d^3 x \; \sqrt{g_{22}g_{33}}\;\rho_{FS} = M\;$, as explained in some discussions about BH mechanics \cite{fourlaws,ashbadri}. The latter option entails  adding a second Lagrange-multiplier term of the form (initially expressed as a scalar, not a density)
\be
\Delta{L}\;=\; \lambda_2 \left(\int_0^R \left[dr ~4 \pi r^2~  \rho_{FS}\right]\;-\;M\right)\;.
\ee
Using Eq.~(\ref{singlepeak}), we may rewrite the constraint term as
\be
\Delta{ L}\;=\; \lambda_2 \left( \int_0^R \left[dr~ 4 \pi r^2 ~E_1 D^1\right]  \;-\;M \right)\;,
\ee
which leads to the following Lagrangian density $\Delta{\cal L}$:
\be
\Delta{\cal L}\;=\; \lambda_2  \frac{\delta\left(r-R\right)}{4\pi r^2} \left(\int_0^R \left[dr~ 4 \pi r^2 ~E_1 D^1\right]  \;-\;M \right)\;.
\ee

We have, for simplicity, assumed a static and spherically symmetric solution in expressing the above constraint term, as this will allow us to show explicitly how the electric force between the charges at the core and the outer surface is canceled by the mass constraint.  However, it is possible to express the mass constraint covariantly, along the lines of  discussions on isolated and dynamical horizons such as \cite{ashbadri}. The outer  layer of the frozen star can be treated as a marginally trapped surface and, therefore, the formalism for calculating the mass as a covariant surface integral is applicable. We leave this to a future investigation.

We can solve for $\lambda_2$ by  varying $\Delta{\cal L}$ with respect to the original gauge field $A^{0}(R)$, as described in  the previous section.~\footnote{Note: The integrand factor of $\;4\pi r^2=4\pi\sqrt{-g}\;$ is unaffected by a covariant derivative.} The equation of motion that results from varying by $A^0(R)$ at the outer surface, $\;\frac{\delta \Delta{\cal L}}{\delta A^0(R)}+J_0(R)=0\;$, along with the Gauss'-law constraint~(\ref{sourceD}) and the relation between the charges~(\ref{J0Out}),  leads to $\;\lambda_2=-1\;$, which  simply means  that the constraint force exactly cancels the  attractive electric force on the outer charge distribution.

\section{Summary and outlook}

We have shown how our frozen star model for an ultracompact BH mimicker can be  described effectively as the
spherically  symmetric solution of Einstein's gravity coupled to the GHY  form of Born--Infeld Lagrangian, which uses Sen's effective action for $D$-brane decay as its starting point.

The current framework can be extended in several directions.
Including rotation is straightforward. Let us recall the zero-angular-momentum-observer's form of the stress tensor for the rotating star, as presented in \cite{notstevekerr}. In this case $\;p=-\rho\;$ and $\;p_{\perp}\sim{\cal O}[\varepsilon^2]\;$, so   that  the corresponding  axially symmetric solution can be found with only an electric field  and vanishing magnetic fields. The only difference from the static, spherically symmetric case is the dependence of the electric field on $r$ and $\theta$ rather than on $r$ alone.

To connect this framework  with the defrosted star, for which $\;\rho+p\;$ and $p_{\perp}$ are no longer vanishing but are perturbatively small, will require  more work. In addition to a spherical electric field, a weak spherical magnetic field will be required.   This can be realized by adding a  magnetic monopole source, meaning that the star is sourced by a dyon, as both its electric and magnetic fluxes are  in the radial direction.

It is also of interest that the Born--Infeld model can be thought of as describing a fluid of open strings, given that the frozen star model has its conceptual origins as a classical version of our earlier-proposed polymer model, which
contains an extremely  hot fluid of closed strings. It is tempting to conjecture that there is
some sort of duality at work here. Nevertheless, of the pair, it is the Born--Infeld description that has an emphatic advantage; namely, a well-defined  mathematical framework, which we look forward to exploiting.

\section*{Acknowledgments}
We thank  Yoav Zigdon for discussions, Suvendu Giri for pointing to us the similarity of the frozen star model to the clouds of string model and to the Born-Infeld BIons and Frans Pretorius for insisting that we find the source matter Lagrangian of the frozen star. We extend our special thanks to Piljin Yi, for helping us understand his work and its implications to the frozen star model. The research is supported by the German Research Foundation through a German-Israeli Project Cooperation (DIP) grant ``Holography and the Swampland'' and by VATAT (Israel planning and budgeting committee) grant for supporting theoretical high energy physics. The research of AJMM received support from an NRF Evaluation and Rating
Grant 119411 and a Rhodes  Discretionary Grant SD07/2022.
AJMM thanks Ben Gurion University for their hospitality during past visits.

\end{document}